\PassOptionsToPackage{shortlabels}{enumitem}
\PassOptionsToPackage{table,xcdraw}{xcolor}

\documentclass[11pt, a4paper]{include/gdm_format}

\usepackage[round]{natbib}
\usepackage{enumitem}
\usepackage[font=small]{caption}
\usepackage{mhchem}
\usepackage{hyperref}
\usepackage{tcolorbox}
\usepackage{xurl}

\definecolor{codeblue}{rgb}{0,0,1}
\definecolor{codegreen}{rgb}{0,0.6,0}
\definecolor{codegray}{rgb}{0.5,0.5,0.5}
\definecolor{codepurple}{rgb}{0.58,0,0.82}
\definecolor{backcolour}{rgb}{0.95,0.95,0.92}
\definecolor{nocolor}{rgb}{1,1,1}

\definecolor{red}{rgb}{0.6,0,0} 
\definecolor{blue}{rgb}{0,0,0.6}
\definecolor{green}{rgb}{0,0.8,0}
\definecolor{cyan}{rgb}{0.0,0.6,0.6}
\definecolor{lightgray}{gray}{0.98}
\definecolor{lightblue}{rgb}{0.13, 0.67, 0.8}
\definecolor{lightorange}{RGB}{255,247,230}
\definecolor{codegreen}{rgb}{0,0.6,0}
\definecolor{codegray}{rgb}{0.5,0.5,0.5}
\definecolor{codepurple}{rgb}{0.58,0,0.82}
\definecolor{keywordcolor}{RGB}{94,20,64}
\definecolor{bluekeywords}{rgb}{0,0,1}
\definecolor{greencomments}{rgb}{0,0.5,0}
\definecolor{redstrings}{rgb}{0.64,0.08,0.08}
\definecolor{xmlcomments}{rgb}{0.5,0.5,0.5}
\definecolor{types}{rgb}{0.17,0.57,0.68}
\definecolor{KWColor}{RGB}{0,0,255}
\definecolor{AnnotationColor}{RGB}{0,137,180}
\definecolor{BlackColor}{RGB}{0,0,0}
\definecolor{CommentColor}{rgb}{0.12,0.38,0.18}
\definecolor{StringColor}{rgb}{0.06,0.10,0.98}
\definecolor{darkred}{rgb}{0.65,0,0}
\definecolor{lightgrey}{rgb}{0.8,0.8,0.8}
\definecolor{marmalade}{RGB}{193,101,18}
\definecolor{peach}{RGB}{250,217,193}
\definecolor{lime}{RGB}{220,237,193}
\makeatletter
\def\@copyrightspace{\relax}
\makeatother

\AtBeginDocument{%
  }

\newcounter{defn}[section]
\renewcommand{\thedefn}{\arabic{section}.\arabic{defn}}

\usepackage{booktabs}
\usepackage{tabularx}
\usepackage{graphicx}

\newcommand{\rev}[1]{{\color{orange}{#1}}}

\begin{document}

\title{Experiment-as-Code Labs: A Declarative Stack for AI-Driven Scientific Discovery}  
%\title{A Declarative Experiment Stack for Autonomous Scientific Labs}
% \title{An Experiment-as-Code Stack for Autonomous Scientific Labs enabling Declarative Experiments}

% \author{University of Michigan\vspace{2mm}}

% Suppress authblk's superscript markers since all authors share one affiliation
\makeatletter
\renewcommand\AB@authnote[1]{}
\renewcommand\AB@affilnote[1]{}
\makeatother

\author{Zhenning Yang}
\author{Yuhan Chen}
\author{Patrick Tser Jern Kon}
\author{Tongyuan Miao}
\author{Hongyi Lin}
% \author{Venkat Viswanathan}
\author{Venkatasubramanian Viswanathan}
\author{Danai Koutra}
\author{Ang Chen}
% \affil{University of Michigan \protect\\ \texttt{\{znyang, cyhalek, patkon, tymiao, lithium, venkvis, dkoutra, chenang\}@umich.edu}}
\affil{University of Michigan}

% \if 0
\begin{abstract} 
To unleash the full potential of AI for Science, we must untether the agents from a purely digital environment. The agent's ability to control and explore in real-world labs is essential because the physical lab remains foundational to scientific discovery. 
While some tasks can be performed on a computer (e.g., data analysis, running simulated experiments), Eureka moments could occur at any time while operating lab instruments (e.g., when a scientist notices unexpected clues, intuition may prompt a real-time course change). 
Although autonomous labs are on the rise, which expose programmable APIs to control scientific instruments via software, bridging the gap between increasingly powerful AI agents and automated lab equipment requires innovation that draws insights from computer systems. 

We propose a new paradigm called ``Experiment-as-Code (EaC) Labs,'' where a core concept is to encode experiments as declarative configurations that can be compiled down to device-level APIs. 
AI agents come up with hypotheses and experiments, written as an ensemble of declarative configurations. 
The systems layer performs program analysis, safety checks, resource assignment, and job orchestration.  
Finally, programmatic experimentation occurs via actuating the device APIs. 
This is a general stack that is science-, lab-, and instrument-independent, representing a novel synthesis across the physical, systems, and intelligence layers to unleash the next breakthrough in AI for Science. 

\end{abstract}

% \fi 

% \author{Paper \#207, 6 pages + references}
%\renewcommand{\shortauthors}{}

\maketitle

\begin{tcolorbox}[
  colback=purple!8!white,
  colframe=purple!8!white,
  boxrule=0pt,
  left=1pt,
  right=1pt,
  top=2pt,
  bottom=2pt
]
Scientific AI agents cannot realize their full potential while confined to digital reasoning.
As autonomous laboratories become shared execution infrastructure for AI-driven discovery, the key bottleneck is no longer only agent intelligence, but the absence of systems abstractions for managing heterogeneous, stateful, and safety-constrained physical instruments.
We argue that the next frontier of ``AI for Science'' is a declarative laboratory stack, driven by AI agents, that compiles experimental intent into safe, state-aware, and reproducible execution across physical laboratories.
\end{tcolorbox}

\section{Introduction}
\label{sec:intro}

AI for Science has seen substantial progress, from AlphaFold~\citep{jumper_highly_2021, Abramson2024-ck} achieving near–experimental accuracy in protein structure prediction, to advances in materials discovery~\citep{materials_2023, materials_2024, materials_2025}, drug design~\citep{drug_2025}, and molecular modeling~\citep{molecular_2020}.
This prompted a recent flurry of proposals from both computer scientists and domain experts advocating an agentic approach to discovery: e.g., using AI agents to propose candidate hypotheses~\citep{hypothesis3, gu2024generation}, run simulations~\citep{ma2024llm}, and analyze data~\citep{chen2024scienceagentbench}.

\looseness=-1
This exciting vision, however, remains hindered by a ``last mile problem'': the majority of existing work studies discovery purely \emph{in silico} (i.e., within the computational domain), but stops short of integrating \emph{physical laboratories} where claims are ultimately established.
Testing any hypothesis, proposed by a human or an AI scientist, eventually requires operating real instruments (e.g., assays, sensors, fabrication) to execute rigorously controlled experiments, measuring outcomes under physical constraints and noise (e.g., calibration drift), and validating that results are reproducible across repeated trials and independent lab setups.
In practice, the path from computational discovery to experimentally validated and deployable technologies typically spans a decade or more~\citep{maine_accelerating_2016}.

Hence, \textit{the frontier of ``AI for Science'' lies in physical lab automation.} To unleash its full potential, we must untether the agent from a purely digital environment. The agent's ability to control and explore in real-world labs is essential. While some tasks may be performed on a computer, Eureka moments can happen at any time when operating lab instruments---e.g., when a scientist notices unexpected clues, intuition may prompt a real-time course change.
Enabling the AI agent to reach the physical lab, therefore, will be fundamental to future breakthroughs.

Agentic control of the scientific lab rests on the premise of programmable control---a trend which has already gained rapid traction and is often dubbed as ``autonomous labs.'' In these labs, scientific instruments (e.g., robotic platforms, synthesis tools) expose programmable APIs for software control (e.g., Python/XML scripts) for high degrees of automation. At the University of Michigan (UM), we have built such labs~\citep{dave_autonomous_2022, chen_elyteos_2025} for battery experimentation; elsewhere, similar labs are found in Emerald Cloud Lab (ECL)~\citep{emeraldcloudlabEmeraldCloud_lab} and a range of other universities~\citep{vriza_self-driving_2023, tom_self-driving_2024, kan_accelerated_2024, kalinin_machine_2023, adams_human---loop_2024, szymanski_autonomous_2023, gao_autonomous_2022, ament_autonomous_2021, liang_real-time_2025, warren_computational_2025, ferreira_da_silva_shaping_2024, lara-ceniceros_advances_2025} and industry companies in the US
and around the world~\citep{seifrid_autonomous_2022, tom_self-driving_2024, song_multiagent-driven_2025, zaki_self-driving_2025, zhang_advancing_2025, starkholm_accelerated_2023, miret_perspective_2025, laws_autonomous_2024, cooper_accelerating_2025, burger_mobile_2020, sanin_integrating_2025}.
Recent startups such as Lila Sciences~\citep{lilaPioneeringScientific_lab} and Periodic Labs~\citep{periodicPeriodicLabs_lab} share our vision of agentic control of the physical lab.

The confluence of the wide variety of autonomous labs, coupled with the evermore powerful AI agents, makes the timing ripe for the idea laid out in this paper.
Such a need is universal across labs (e.g., UM vs.\ ECL), instruments (e.g., cyclers vs.\ pumps), and sciences (biology vs.\ materials), motivating a general stack that enables AI agents to operate physical instruments safely, reliably, and with high reproducibility, while unleashing AI models' intelligence. Building such a \textit{scientific stack} in a lab-, instrument-, and science-independent manner is a grand challenge: at its core, we need computer systems support that enables ``AI for Science.''

\begin{figure}[t]
\centering
\includegraphics[width=.99\linewidth]{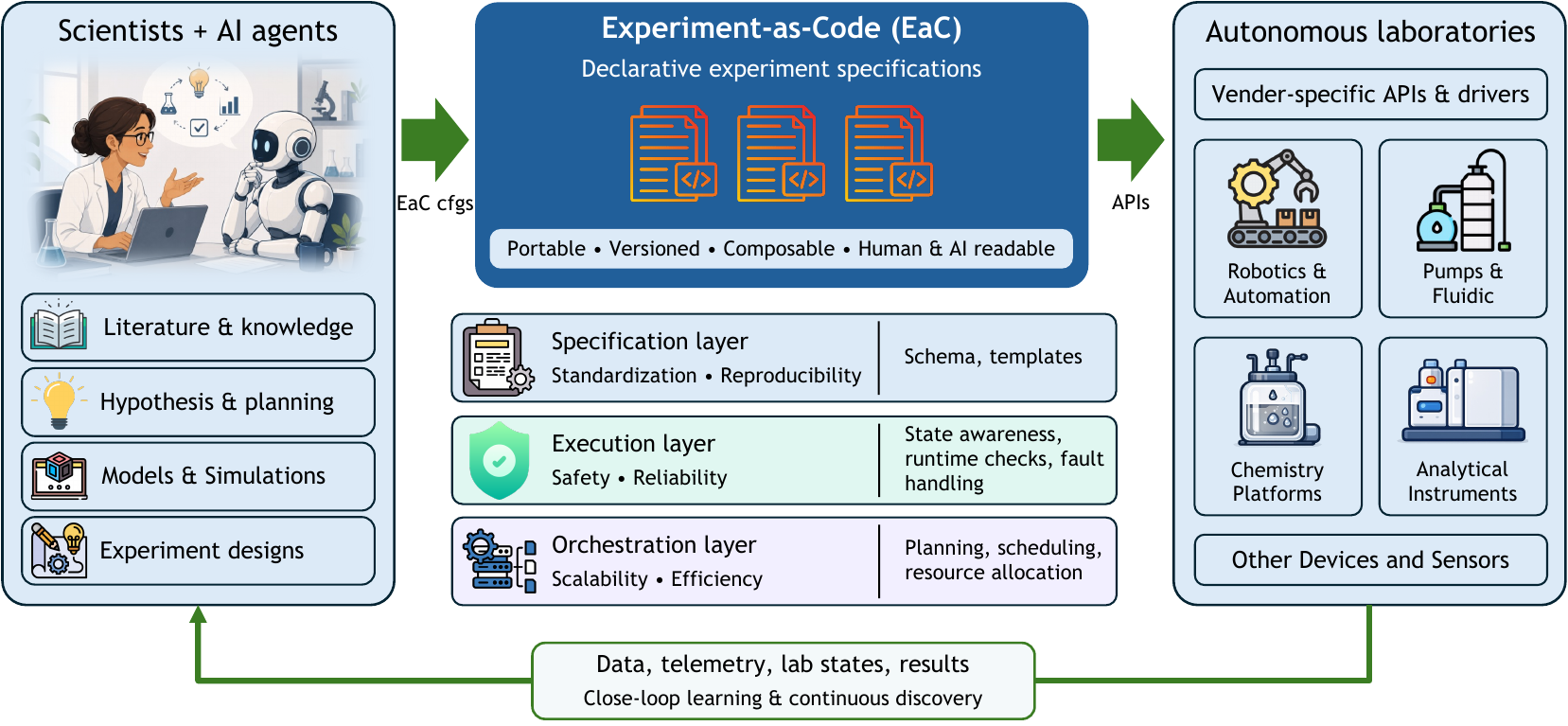}
\caption{Experiment-as-Code (EaC): a declarative stack that bridges AI-driven scientific reasoning and autonomous laboratory execution.}
\label{fig:overview}
\vspace{3mm}
\end{figure}

We propose such a stack in a new paradigm called Declarative Experiment Stack, shown in Figure~\ref{fig:overview}, where a core concept is to encode experiments as declarative configurations, which we call Experiment-as-Code (EaC) programs, that can be compiled down to instrument-level APIs.
Everything in the scientific lab is controlled and optimized via this programmable declarative experiment layer. AI agents come up with hypotheses and experiments, which are written as an ensemble of declarative EaC configurations (e.g., parameter sweeping experiments). The stack performs EaC code analysis to check for safety, calibrate instrument drift, track provenance, assign resources, and schedule the jobs for execution.
New and heterogeneous instruments can be on-boarded to our stack via a shim layer that encodes these new entities in the EaC schema.
EaC programs are portable across labs for high reproducibility.
This new paradigm enables a novel synthesis across the physical, systems, and intelligence layers to transform the scientific lab for digital control.

\begin{figure}[t]
\centering
\includegraphics[width=.95\linewidth]{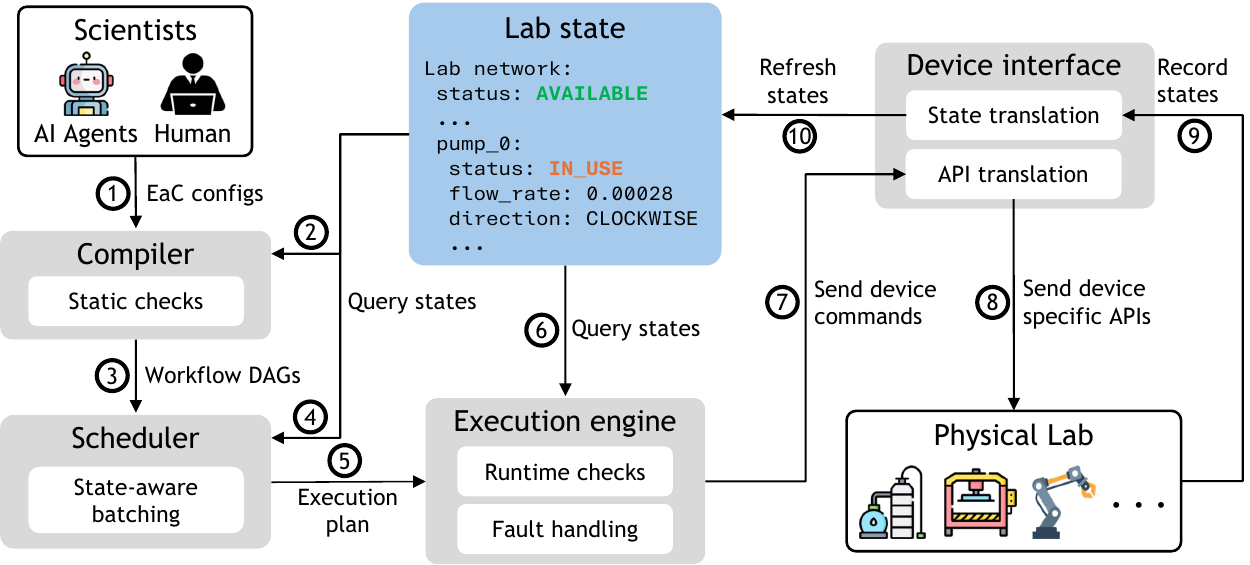}
\caption{The EaC framework bridges declarative experiment specifications and physical lab automation. EaC configs authored by scientists (1) are statically checked by the compiler against the lab state (2) and lowered into workflow DAGs (3). The scheduler consults lab state (4) to produce a state-aware execution plan (5), which the execution engine dispatches with runtime checks against live state (6, 7). The device interface translates unified operations into vendor-specific API calls (8), records returning telemetry (9), and refreshes the centralized lab state (10).}
\label{fig:workflow}
\vspace{3mm}
\end{figure}

This design is inspired by Infrastructure-as-Code (IaC)~\citep{terraform, qiu_simplifying_2023}, which has been instrumental to modern cloud infrastructure operations.
Transporting and adapting these principles to the scientific lab enables a range of benefits. It accelerates discovery by automating routine procedures reliably; improves reproducibility by treating experiment definitions as versioned code; reduces errors through static validation before execution; enables direct integration with AI agents through well-defined digital interfaces; and supports cloud-like sharing and utilization of laboratory resources.
% Figure~\ref{fig:workflow} presents EaC as the interface layer between human/AI reasoning and physical lab equipment.
Figure~\ref{fig:workflow} presents the core EaC architecture linking human/AI reasoning to physical laboratory execution.
Human scientists and AI agents specify experiments through EaC interfaces, while EaC compiles, schedules, and executes workflows against a centralized, continuously maintained lab state.
Device telemetry updates this state, enabling faster closed-loop iteration between experiment design and lab execution.

To ground our approach, we applied the EaC design to an existing autonomous battery laboratory at the University of Michigan and executed real experimental workflows end-to-end.
We present this case study at Sec~\ref{sec:um-case-study}.
Previously, experiments were implemented as device-specific, imperative scripts that tightly coupled scientific intent with low-level hardware control.
Writing and modifying these procedures required substantial expertise, and the resulting code was difficult to interpret, reproduce, or extend.
With EaC, experiments are expressed as concise, declarative specifications that describe formulations and measurements without embedding device logic.
The framework validates device states, enforces safety checks, and compiles specifications into coordinated API calls across laboratory equipment, substantially reducing human effort.

\noindent\textbf{Roadmap and contributions.}
% The remainder of the paper unfolds the EaC vision in five steps.
The remainder of the paper unfolds the EaC vision as follows.
Section~\ref{sec:related-work} situates EaC at the intersection of two parallel threads (AI scientists and autonomous laboratories), and argues that neither thread alone closes the gap between digital reasoning and physical execution.
It also draws inspiration from cloud computing's transition from imperative consoles to declarative infrastructure-as-code, and motivates an analogous declarative layer for shared, stateful labs. 
Section~\ref{sec:eac-core} introduces EaC's three-layer stack (specification, execution, and orchestration), and identifies the desired properties of scientific automation that each layer enables: standardization and reproducibility at the specification layer, safety and reliability at the execution layer, and scalability and efficiency at the orchestration layer.
Section~\ref{sec:um-case-study}  grounds the design in a case study of an existing autonomous battery laboratory at the University of Michigan, where we rearchitect a script-driven platform into an EaC-enabled lab and run a real ionic conductivity experiment end-to-end.
Section~\ref{sec:conclusion} closes with open research questions and a call to action for the systems, AI, and scientific communities. 
Together, these contributions chart a concrete path toward a programmable, AI-native laboratory stack that is portable across laboratories, instruments, and scientific domains. 
 
\section{Related Work}
\label{sec:related-work}

Prior work on automating scientific workflows largely falls into two parallel threads: \emph{AI scientists} that use large models to reason about hypotheses, experimental design, and data analysis, and laboratory automation systems that execute experiments through specialized software and hardware infrastructure.
While both aim to accelerate scientific discovery, they operate at different abstraction levels: AI scientists primarily reason in digital environments without physically executing plans on the laboratory instrument, whereas automation systems focus on execution but encode experiments in ad hoc, non-portable forms.
This disconnect motivates the need for a unified, program-centric abstraction that supports both reasoning and execution of experiments in real laboratory settings.

\noindent
\textbf{AI scientists.}
Recent work on AI for Science investigates LLM-powered agents that read literature, propose hypotheses, plan experiments, and analyze results, often described as \emph{AI Scientists}~\citep{tie2026surveyaiscientists}.
Existing works have explored the capabilities of these LLM agents across the scientific workflow.
These agents can conduct literature review~\citep{lala2023paperqa_liter_review2, agarwal2024litllm_liter_review3, schmidgall2025agentlaboratoryusingllm_liter_review1} and contextualization, generate and iteratively refine research ideas~\citep{Zhou_2024_idea1, su-etal-2025-many_idea2, vasu-etal-2025-hyper_idea3, ghafarollahi2025sciagents_idea4}.
Designing and executing experiments \textit{in silico} or on automated platforms~\citep{kon_curie_2025, boiko2023autonomous_exp1, mandal2025evaluating_exp3, kon_exp-bench_2025}.
Analyzing data and synthesizing findings, drafting papers, and simulating peer review~\citep{zhou2025largelanguagemodelspenetration_review1, lala2023paperqa_liter_review2, agarwal2024litllm_liter_review3, li2024exploring_writing1, cheng2025artificial_writing2, khalifa2024using_writing3}.
A smaller subset of recent ``AI scientist'' frameworks integrates several of these capabilities into end-to-end agentic pipelines, aiming to autonomously close the loop from problem formulation to paper writing~\citep{tang2025airesearcherautonomousscientificinnovation_e2e1, yamada2025aiscientistv2workshoplevelautomated_e2e2, lu2024aiscientistfullyautomated_e2e3}.
However, these systems are predominantly developed and evaluated within computer science (CS) settings, where CS experiments are digitally executed, reproduced, and validated.
In contrast, experimental workflows in domains such as chemistry and biology involve physical devices, safety constraints, and irreversible actions, making verification a critical and labor-intensive step that still relies heavily on human expertise.
While AI Scientists can suggest experimental designs, translating them into executable laboratory procedures and validating their correctness and feasibility typically requires substantial manual effort.

\noindent
\textbf{Autonomous laboratories.}
Existing laboratory automation software, such as LIMS and ELNs, can automate individual components of experimental workflows, including sample tracking, instrument control, data capture, and scheduling, to improve efficiency and traceability~\citep{materialszone_lab_automation_2024,lablynx_lab_software_overview_2025,starlims_eln_lims_integration_2025}.
However, these systems typically do not support end-to-end experimental execution.
Moreover, experiments are commonly encoded as GUI-driven workflows or vendor-specific configuration templates that are tightly coupled to a single platform's instrument library and data schema. 
While these representations expose a declarative interface at the surface level, they are not portable, statically checkable, or composable across labs: they describe \emph{which buttons to click in which tool} rather than \emph{what experiment to perform}, leading to siloed ecosystems that evolve slowly and continue to require substantial human effort for experimental implementation, verification, and day-to-day operation~\citep{materialszone_lab_automation_2024,scispot_top_lims_2026,holland_automation_life_science_2020}.
In contrast, EaC envisions a well-typed, lab-independent program over a standardized resource model, with explicit dependencies, capability schemas, and execution semantics that admit static validation, deterministic compilation, and cross-site execution. 
These limitations have motivated the emergence of autonomous and ``cloud'' laboratories (e.g., Emerald Cloud Lab~\citep{emeraldcloudlabEmeraldCloud_lab}), which provide remotely accessible, highly automated facilities that allow researchers to configure, execute, and analyze experiments through centralized software interfaces without being physically present.
By offering on-demand access to specialized and expensive equipment, these platforms draw analogies to cloud computing infrastructures in the goal of lowering the barrier to conducting complex experiments~\citep{arias_cloud_labs_2024}.
However, as the number of instruments and analytical capabilities grows, configuring experiments increasingly resembles orchestrating many cloud services, imposing steep learning curves and exposing users to imperative, low-level interfaces that remain cumbersome and error-prone, ultimately limiting the efficiency gains of laboratory automation in practice.

\noindent \textbf{National and international initiatives.} Recent policy frameworks worldwide have advocated the integration of AI with scientific infrastructure, recognizing automation as a strategic imperative for accelerating discovery. In the United States, the National Science Foundation has launched a \$100 million initiative to build a national network of AI-programmable cloud laboratories~\citep{nsf_2025}, aiming to democratize access to automated experimentation via the PCL testbed~\citep{nsf_test_bed_2025}. Simultaneously, the White House's Genesis mission directs the Department of Energy to build an integrated AI-for-science platform that couples high-performance computing with autonomous experimental facilities~\citep{genisis_2025}. Similar efforts are gaining momentum globally: the European Union's ``RAISE'' strategy envisions a virtual institute to coordinate AI resources, compute, data, and scientific talent~\citep{raise_bill_2025, eu_raise}, while China's ``AI+'' action plan explicitly targets the intelligent upgrading of major science infrastructure and the promotion of new AI-driven research paradigms~\citep{china_aip, china_aip_en}. Collectively, these initiatives signal a growing consensus that the future of science depends on closing the loop between digital intelligence and physical execution.
DEX provides a concrete technical roadmap for realizing this vision by introducing a unified, declarative experiment layer that bridges AI reasoning with safe, reproducible, and portable execution in heterogeneous physical laboratories.

%\rev{\section{Lessons from Cloud Computing}}
%\label{sec:cloud-lessons}

%\zy{this still seems a bit sudden. To this point, it's not clear why it is related to cloud computing. consider hint it earlier, maybe even have remote or cloud in the title?}

\begin{table}[t]
\centering
\resizebox{\linewidth}{!}{%
\begin{tabular}{c|cc}
\toprule
\textbf{Interface Model} & \textbf{Cloud Computing} & \textbf{Autonomous Laboratories} \\
\midrule
Imperative
& Web consoles
& Web consoles \citep{emeraldcloudlabEmeraldCloud_lab} \\

Imperative
& APIs/CLI/SDK
& APIs/Wolfram Language~\citep{emeraldcloudlabEmeraldCloud_lab, Wolfram} \\

Declarative
& IaC \citep{terraform}
& \textbf{Experiment-as-Code} \\

\bottomrule
\end{tabular}%
}
\caption{Analogies between cloud computing platforms and autonomous laboratories: the evolution from imperative to declarative interfaces improves manageability.}
\label{tab:interface_evolution}
\vspace{3mm}
\end{table}

\noindent \textbf{Cloud computing.}
We see many analogies to cloud computing. Autonomous laboratories today look strikingly similar to cloud platforms: a heterogeneous fleet of resources, accessed remotely, used by multiple tenants, and operated through cloud-specific APIs. 
A recent trend that simplifies cloud management is Infrastructure-as-Code (IaC), which exposes a provider-independent interface for programming cloud services.
Users declare the \emph{desired state} of their system, and a compiler translates that intent into low-level execution plans~\citep{terraform, qiu_simplifying_2023}. 
Similar principles have been explored in prior work~\citep{Aguilar2024-mh}, which combines existing tools such as Jupyter Notebook~\citep{jupyter}, Docker~\citep{docker}, and Terraform~\citep{terraform} with documentation to support reproducible VR experimental workflows.
We draw inspiration from these successes to separate intent from execution, although EaC must contend with physical realities that the cloud can largely abstract away: instruments are stateful and not cleanly time-sliceable, calibration drifts, safety constraints depend on live physical conditions, and many primitives are irreversible (e.g., consumed reagents, contaminated wells).
Our design borrows from cloud principles where they transfer cleanly and rethinks them where physical execution demands new abstractions.
Table~\ref{tab:interface_evolution} summarizes this contrast.

\section{Experiment-as-Code Labs}
\label{sec:eac-core}

\begin{figure}[t]
\centering
\includegraphics[width=.99\linewidth]{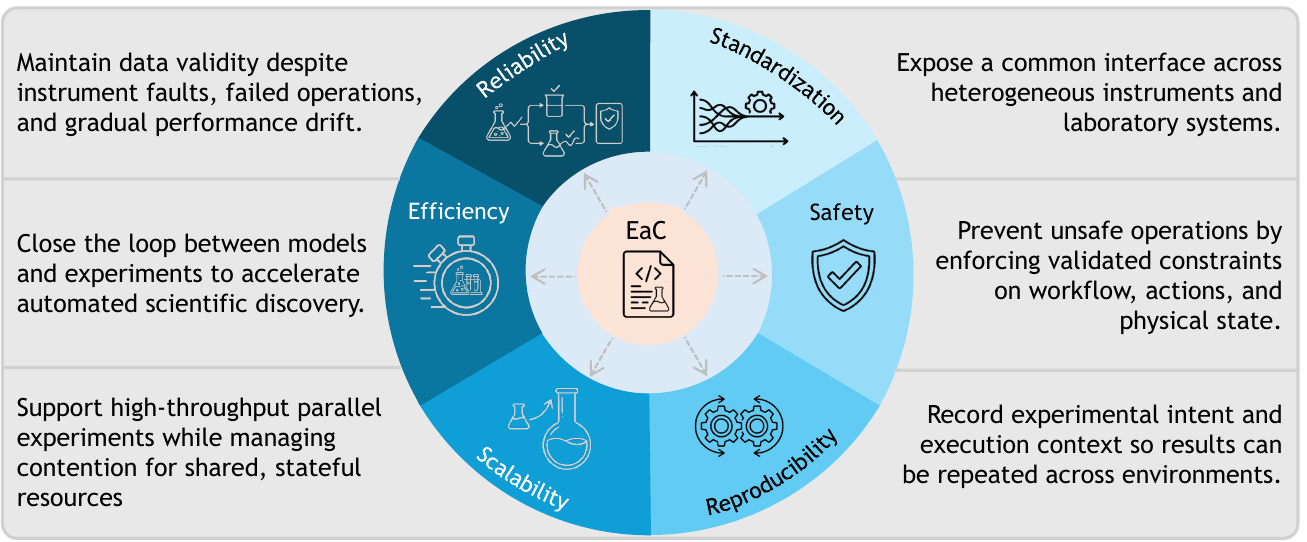}
% \captionsetup{justification=centering}
\caption{Key benefits of Experiment-as-Code labs. EaC translates scientific intent into standardized and reproducible specifications, enables safe and reliable execution, and supports scalable, efficient orchestration across shared, stateful laboratory resources.}
\label{fig:pillars}
\vspace{3mm}
\end{figure}

Experiment-as-Code laboratories turn experiments into programmable artifacts that can be version-controlled and readily replicated across sites, addressing longstanding challenges of reproducibility and error-prone manual setups. 
Encoding procedures as code also enables static checks and automated validation, catching unsafe or incompatible configurations before they run. 
The EaC framework organizes the stack in three layers. 
Figure~\ref{fig:pillars} summarizes the key benefits enabled by these layers.
\begin{itemize}[leftmargin=*,itemsep=1pt,topsep=0pt,parsep=0pt]
    \item \textbf{Specification layer.} EaC expresses experimental intent as portable, declarative configurations over standardized device interfaces. This layer supports \textit{standardization} by hiding vendor-specific APIs behind unified abstractions, and \textit{reproducibility} by making resources, parameters, dependencies, and execution semantics explicit.
    
    \item \textbf{Execution layer.} EaC validates specifications against device capabilities, safety constraints, calibration assumptions, and live laboratory state, then enforces these constraints during physical operation. This layer supports \textit{safety} by preventing invalid or hazardous actions, and \textit{reliability} by detecting device faults, drift, and other state violations during execution.
    
    \item \textbf{Orchestration layer.} EaC compiles validated specifications into executable workflow DAGs and schedules them over shared, stateful instruments. This layer supports \textit{scalability} by managing resource contention and parallel execution, and \textit{efficiency} by reducing redundant work, unnecessary waiting, and manual translation from scientific intent to device-specific procedures.
\end{itemize}

\noindent The declarative experiment stack aims to reduce the cognitive and operational burden on scientists---especially for AI agents who are well-versed at programmable tasks. By offloading low-level coordination, checks, and constraints, EaC allows agents to focus on creative hypothesis generation and experimental design while maintaining rigor.
Next, we present each layer by outlining the limitations of the current design, deriving a systems observation, and describing the corresponding EaC design.

\subsection{Specification Layer}
\label{subsec:specification}

The specification layer defines how experimental intent is represented before execution, turning heterogeneous device operations and implicit procedural assumptions into explicit, portable experiment descriptions.
This layer addresses two related limitations in existing automation: device interfaces are not standardized, and execution semantics needed for reproducibility are often implicit.

\subsubsection{Standardization}
\label{subsubsec:standardization}

\begin{table*}[t]
\centering
\resizebox{0.9\textwidth}{!}{

\begin{tabular}{l p{9cm} p{5cm}}
\toprule
\textbf{Device} & \textbf{Low-Level Device Control} & \textbf{EaC Abstraction} \\
\midrule

Pump &
\begin{minipage}[t]{9cm}\ttfamily
cmd = bytearray([0xE9,0x0E,0x08,...])\\
ser.reset\_input\_buffer()\\
ser.write(cmd)\\
ser.flush()
\end{minipage}
&
\begin{minipage}[t]{5cm}\ttfamily
pump\_2.dispense:\\
\{flow\_rate:4,volume:0.7\}
\end{minipage}
\\

\midrule

Valve &
\begin{minipage}[t]{9cm}\ttfamily
cmd = b"G005\textbackslash r"\\
ser.write(cmd)\\
ser.flush()
\end{minipage}
&
\begin{minipage}[t]{5cm}\ttfamily
valve.set:\{dest:5\}
\end{minipage}
\\

\midrule

Balance &
\begin{minipage}[t]{9cm}\ttfamily
line = ser.readline().decode(...)...\\
match = re.search(r'\textbackslash d+\textbackslash.\textbackslash d+', line)\\
mass = float(match.group())\\
is\_stable = ('?' not in line)
\end{minipage}
&
\begin{minipage}[t]{5cm}\ttfamily
balance.read:\{\}
\end{minipage}
\\

\midrule

Relay &
\begin{minipage}[t]{9cm}\ttfamily
device = hid.device()\\
device.open\_path(PATH)\\
cmd = [0x00,0xFF,channel]\\
device.send\_feature\_report(cmd)
\end{minipage}
&
\begin{minipage}[t]{5cm}\ttfamily
relay\_1.on:[1]
\end{minipage}
\\

\midrule

% Potentiostat (assembly-level) &
Potentiostat &
\begin{minipage}[t]{9cm}\ttfamily
mov rcx, hSerial\\
lea rdx, buffer\\
mov r8d, length\\
lea r9, bytesWritten\\
mov qword ptr [rsp+32],0\\
call WriteFile
\end{minipage}
&
\begin{minipage}[t]{5cm}\ttfamily
palmsens4.configure:\\
\{eac:0.25, freq\_min:20000, freq\_max:592000, n\_freq:10\}
\end{minipage}
\\

\bottomrule
\end{tabular}

}
% \caption{Comparison between low-level device control and Experiment-as-Code (EaC) abstraction.}
\caption{Examples of device-specific control code and the corresponding EaC abstractions. EaC replaces low-level byte commands, serial I/O, parsing logic, and vendor-specific APIs with compact, typed operations that expose scientific intent while hiding device implementation details.}
\label{tab:eac-abstraction}
\vspace{3mm}
\end{table*}

Laboratory instruments expose diverse vendor APIs with inconsistent command interfaces, telemetry formats, and safety semantics, making existing automation brittle and device-specific. 
% \rev{
Table~\ref{tab:eac-abstraction} illustrates this gap: simple scientific operations often require low-level byte commands, serial communication, parsing logic, or vendor-specific calls, whereas EaC exposes the same operations through compact, typed abstractions.
% }

\noindent\textbf{Inconsistent command interfaces.} 
Even for common operations, vendors disagree on the \emph{primitive} and its parameters.
For centrifugation, some controllers expose speed in \texttt{rpm} while others accept (or prefer) relative centrifugal force \texttt{rcf/$\times g$}, requiring rotor-dependent conversions and different parameterizations across devices~\citep{eppendorf5910_manual,eppendorf_rpm_rcf, ThermoSciRC6Manual}.
For temperature control, devices may require an explicit \emph{ramp-rate/profile} specification (not just a target setpoint), e.g., thermal cyclers define steps with temperature setpoints \emph{and} ramp rate settings~\citep{thermofisher_atc_manual}.
Even the execution semantics differ: many SCPI-style devices provide a separate ``operation complete'' query that \emph{blocks} until prior commands finish (vs.\ fire-and-forget commands that return immediately), forcing client code to encode device-specific synchronization~\citep{keysight_scpi_opc}.

\noindent\textbf{Inconsistent telemetry formats.}
Some instruments natively produce dense time-series outputs (e.g., kinetic microplate reads return a full absorbance time course per well), while others provide only sparse alarms or coarse status signals.
For instance, microplate software commonly represents kinetic assays as per-well time-series measurements~\citep{biorad_microplate_manager_manual},
whereas some incubators expose ``remote monitoring'' primarily via an alarm relay contact for out-of-range conditions rather than a structured high-rate telemetry stream~\citep{vwr_incubator_manual}.
More generally, industrial telemetry stacks (e.g., OPC UA) often encode health/quality via numeric \texttt{StatusCode}s, which can be opaque without vendor-specific mapping and consistent timestamp/unit conventions~\citep{opcua_statuscode}.

\noindent\textbf{Inconsistent safety semantics.}
Where safety is enforced (inside the device's own firmware versus inside the host software that drives it) differs sharply across instruments, and the boundary is rarely documented.
Some devices treat safety as a hardware-enforced invariant: many centrifuges implement hard interlocks (e.g., preventing operation with the lid open or until safe conditions hold) that the host cannot bypass, no matter what the controlling script issues~\citep{unl_centrifuge_safety}.
Other devices delegate safety enforcement to the host, surfacing unsafe conditions only as alarms or status flags (e.g., relay-triggered alarm states) and relying on the automation client to enforce preconditions and decide whether to pause or abort; here, an omission in client-side checks is itself a safety incident~\citep{vwr_incubator_manual}.
A portable automation framework must therefore reason explicitly about \emph{which side of the boundary} a given safety property lives on for each device, rather than assuming a uniform contract.

\begin{tcolorbox}[colback=blue!5!white,colframe=blue!50!black,left=1pt, right=1pt, top=2pt, bottom=2pt]
\textbf{Observation \#1:} A primary barrier to lab automation is the lack of a shared operational abstraction across devices. In the absence of standardized interfaces and a unified state model, it poses a significant learning curve, requiring developers and users to reason about vendor-specific APIs, parameter conventions, and operational behaviors at a low level.
\end{tcolorbox}

The lack of a shared interface is especially problematic for AI agents.  
Every vendor SDK is a low-resource ``language'' in the LLM sense: training data is scarce, idioms are inconsistent, and parameter conventions (rpm vs.\ rcf, blocking vs.\ fire-and-forget, alarm flags vs.\ structured telemetry) shift between devices that nominally do the same thing.
As a result, an agent that operates correctly in one lab cannot transfer to another without per-lab fine-tuning, prompt engineering against bespoke documentation, or custom tool wrappers, even when the underlying scientific procedure is identical.
This is precisely the gap that a shared, instrument-independent abstraction can close: by mapping heterogeneous vendor APIs onto a small, stable, well-documented programming surface, EaC turns lab control into a standardized programming target that AI agents can learn once and reuse everywhere.

\subsubsection{Reproducibility} 
\label{subsubsec:reproducibility}

Reproducibility in physical experimentation requires more than repeating high-level procedures; it demands that experimental intent be captured in a precise, executable form whose semantics is independent of specific instruments or implementations.
In existing automation frameworks, experimental procedures are often encoded as scripts that directly invoke vendor APIs, implicitly inheriting undocumented assumptions about timing, defaults, and execution order. As a result, even small changes in hardware, firmware, or lab configuration can alter the realized behavior of an experiment with the same script~\citep{Hawkese2555, Baker2016}. 

\noindent\textbf{Semantic mismatch.}
Even seemingly unambiguous steps can diverge at the semantic level because real instruments expose (and often require) hidden “stabilization” and “geometry” details, among others.
For incubation, some CO$_2$ incubator manuals explicitly recommend waiting \emph{after the display reaches setpoint} (e.g., ``allow at least 2 hours after the display reaches setpoint for temperature to stabilize''), so ``incubate for 24 hours'' can mean 24 hours from command issuance vs.\ 24 hours after stabilization---shifting effective exposure time and conditions, as documented in incubators where stabilization itself can take tens of minutes to hours after conditions are set~\citep{thermofisher_stericycle_manual}.
For plate washing, ``wash plate'' is underspecified unless the scientist fixes parameters like aspiration height/depth, dispense height, and whether the step includes shake/soak; washer manuals expose these as explicit protocol primitives whose settings materially affect residual volume and background~\citep{biotek_el406_manual,thermofisher_wellwash_versa_manual}.

\noindent\textbf{Script non-portability.}
Scripts often “work” by implicitly relying on device- and plate-specific defaults that do not transfer.
For instance, washer software uses device-specific plate parameterizations 
(e.g., plate-type tables where aspiration height defaults flow 
from plate selection~\citep{biotek_el406_manual, biotek_405ls_manual}, or systems 
where plate geometry and aspiration height are decoupled 
protocol parameters~\citep{moleculardevices_aquamax_userguide}), so omitting these parameters can yield systematically different washing outcomes across washers even when the high-level logic is identical.
Similarly, incubator control behaviors (e.g., setpoint recovery dynamics and stabilization 
delays) differ across platforms, so ``incubate at 37\,°C'' is sensitive to device-specific defaults and stabilization semantics unless these are made explicit~\citep{thermofisher_stericycle_manual,nuaire_co2_incubator_manual}.

\begin{tcolorbox}[colback=blue!5!white,colframe=blue!50!black,left=1pt, right=1pt, top=2pt, bottom=2pt]
\textbf{Observation \#2:} Reproducibility breaks when critical execution details---such as stabilization delays, geometry-dependent parameters, and device-specific defaults---remain implicit in scripts rather than encoded as explicit experimental intent. As a result, identical procedures can lead to different physical outcomes when executed on different instruments or laboratory configurations.
\end{tcolorbox}

% \textcolor{red}{check AI.} 
% \rev{
Even expert human scientists frequently fail to reproduce published experimental results, as documented in prior studies~\citep{Baker2016, Hawkese2555}. 
This reflects a semantic gap in which publications describe high-level procedures but often omit the execution semantics needed to reproduce experimental results.
These gaps are further exacerbated by heterogeneous instruments and laboratory settings, where nominally identical procedures are instantiated through different interfaces, defaults, and calibration states.
Reproducibility is a major blocker for AI-driven automated discovery. 
Recent work in the AI community~\citep{hypothesis3, Zhou_2024_idea1, vasu-etal-2025-hyper_idea3, HuangKKPWYY0025} has explored scientific hypothesis generation, where agents synthesize new ideas grounded in prior literature. 
This process fundamentally assumes that prior results are reliable and reproducible. 
When this assumption fails, agents may generate large numbers of hypotheses based on underspecified or non-portable experiments, leading to wasted downstream effort and inconsistent outcomes. 
We argue that a reproducibility-preserving abstraction is beneficial to the scientific community and, more importantly, a key enabler for robust and rigorous AI-driven scientific discovery.
% }

\subsubsection{EaC's Specification Layer Addresses these Limitations}

\begin{figure}[t]
\centering
\includegraphics[width=0.99\linewidth]{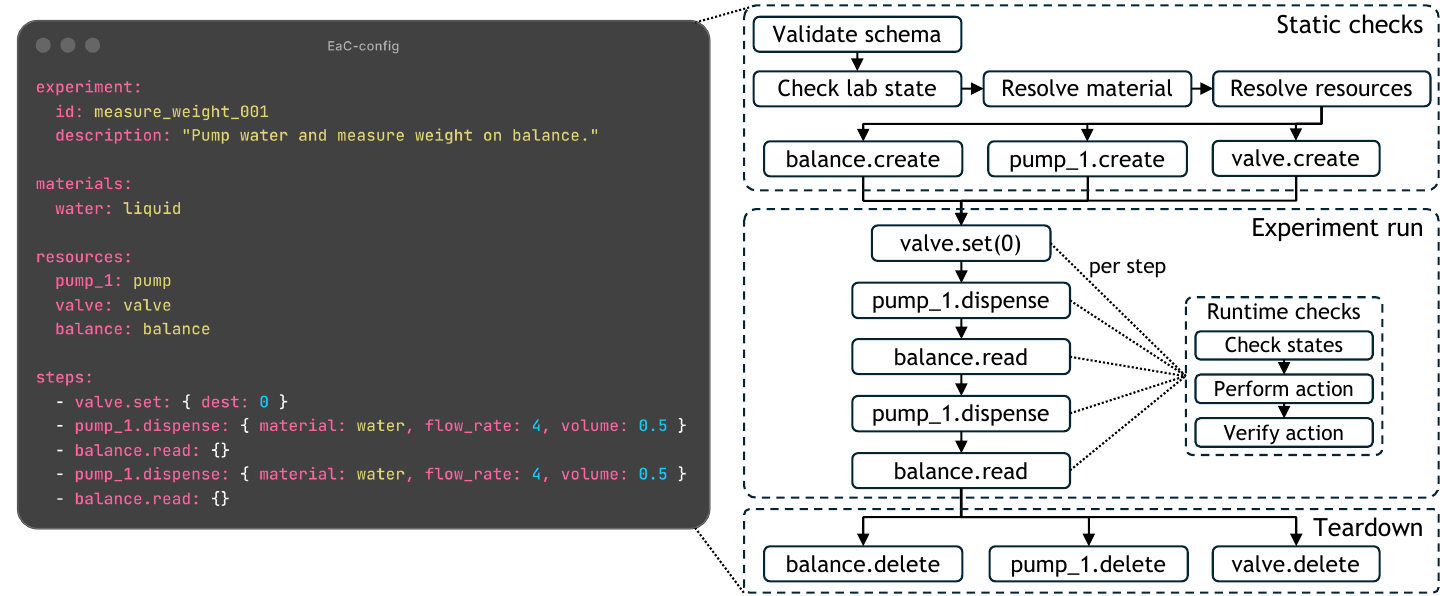}
\caption{Example EaC configuration and its compiled execution graph. Declarative experiment steps are automatically expanded into device-level operations with inferred checks, dependencies, and teardown.}
\label{fig:eac-example}
\vspace{3mm}
\end{figure}

EaC addresses these limitations by separating experimental intent from device-specific execution through a declarative specification layer.
Users, including human scientists and AI agents, describe what experiment should be performed using portable configurations over standardized device resources, rather than writing imperative scripts against vendor-specific APIs.
Each resource exposes a stable set of high-level operations for configuration, actuation, and telemetry.
Device wrappers map these operations to low-level commands and translate vendor-specific status outputs into a unified state schema.
This specification layer makes the assumptions required for reproducible execution explicit.
EaC configurations identify required resources, operation parameters, dependencies, timing constraints, stabilization requirements, and measurement conditions that would otherwise be hidden in scripts or device defaults.
The compiler then resolves each declarative operation into a structured workflow graph whose nodes invoke well-defined resource operations with explicit parameters.
Figure~\ref{fig:eac-example} illustrates this compilation process for an example EaC configuration.
By combining standardized resource interfaces with declarative experiment descriptions, EaC provides a portable representation of experimental intent.
The same specification can be interpreted across heterogeneous instruments through capability schemas and wrapper mappings, while preserving the execution semantics needed for repeatability.
In this way, standardization becomes the mechanism that enables reproducible experiment specification across laboratories.

\subsection{Execution Layer}
\label{subsec:execution}

\looseness=-1
The execution layer governs how validated experiment descriptions are carried out in the physical lab, ensuring that safety constraints, live device state, and failure conditions are checked throughout the operation.
This layer addresses two related risks in physical execution: unsafe actions can arise from missing live-state validation, and unreliable results can arise from explicit faults or silent state violations.

\subsubsection{Safety}
\label{subsubsec:safety}

Safety in physical experimentation requires that a program be executable only when its workflow logic respects operational constraints and the live laboratory state satisfies device-specific preconditions.
Ensuring safety is difficult because hazards can arise from complex, stateful, and sometimes unpredictable instrument behavior.

\noindent\textbf{Context-dependent exposure hazards.}
Safety constraints are not purely mechanical.
For example, biosafety guidance highlights that certain procedures have elevated aerosol risk and require additional containment and controls based on activity-specific risk assessment~\citep{cdc_bmbl6,cdc_monkeypox_biosafety}.
Concretely, public health guidance calls out mitigations such as \emph{sealed centrifuge rotors/safety cups} and appropriate PPE to reduce personnel exposure risk during centrifugation and related handling~\citep{cdc_monkeypox_biosafety}.
These are “invisible” to typical device APIs: the hazard depends on whether the rotor is sealed, what is being spun, and the facility’s biosafety posture—exactly the kind of safety context that must be represented in the lab state and enforced as a precondition.

\noindent\textbf{Transient thermal and mechanical hazards.}
Many plate sealing and heating workflows create transient burn/crush hazards that must be respected during automated and human-in-the-loop execution.
For instance, plate sealers explicitly warn that metal parts/platen surfaces can remain hot after use and should be allowed to cool before handling or maintenance~\citep{biorad_px1_sealer_manual,corning_platemax_sealer_manual}.
Heat-sealer instructions similarly warn that heater plates can reach very high temperatures and remain hot for a considerable time, requiring strict operating precautions~\citep{coleparmer_heat_sealer_ifu}.
Even if the code is correct, the \emph{handoff} between devices (or to a human) can be unsafe unless the runtime enforces cooldown windows, temperature thresholds, and “safe-to-touch/transfer” predicates.

\begin{tcolorbox}[colback=blue!5!white,colframe=blue!50!black,left=1pt, right=1pt, top=2pt, bottom=2pt]
\textbf{Observation \#3:} Many safety risks arise when execution proceeds without validating live device state and operational preconditions. Even correct workflows can become unsafe if containment status, cooldown periods, or calibration constraints are not enforced during execution.
\end{tcolorbox}

\subsubsection{Reliability}
\label{subsubsec:reliability}

Reliability matters in physical experimentation because failures are inevitable, yet the lab must uphold scientific rigor: a robust system should (i) detect failures promptly, (ii) contain blast radius by preventing downstream corruption, and (iii) recover deterministically when possible. 
In practice, failures fall into two complementary classes:
\emph{explicit failures}, where the device or controller reports an error or becomes unavailable, and 
\emph{implicit failures}, where the device continues to operate but silently violates the accuracy or state assumptions required for valid results (e.g., calibration lapse, drift, or invalid comparability conditions).

\noindent\textbf{Explicit failures.}
Some automation steps fail in ways that are \emph{observable}, but are hard to recover from.  
Liquid handlers, for instance, commonly experience runtime exceptions when sensing and aspiration do not behave as expected (e.g., ``no liquid detected'' or ``insufficient liquid detected''), which can arise from empty wells, misaligned labware, or transient detection noise~\citep{tecan_lld_errors}.
Similarly, plate washers and dispensers expose device-level failures through explicit error codes/messages; operator manuals instruct users to record error codes and consult error tables, reflecting that these faults occur during normal operation (e.g., motion/positioning faults, fluidic faults, or initialization failures)~\citep{biotek_el406_manual}.
At the control layer, instruments driven via command/response protocols can also fail due to communication timeouts or termination/handshake mismatches, producing a \emph{clear} failure signal but leaving ad hoc scripts uncertain whether a command executed, partially executed, or did not execute at all~\citep{keysight_3458a_timeout}.
Without a principled failure model (retry/abort semantics, idempotence boundaries, and state reconciliation), these explicit faults often lead to either unrecoverable aborts or unsafe “best-effort” continuation.

\noindent\textbf{Implicit failures.}
More dangerous are failures that do not raise immediate errors, but invalidate the assumptions needed for correctness.
A canonical case is sensor calibration validity: CO$_2$ incubators and similar controlled-environment instruments require periodic verification/calibration of temperature/CO$_2$/humidity sensors, with vendor guidance commonly recommending calibration upon installation and regularly thereafter (e.g., annually)~\citep{baker_reco2ver_calibration}.
If calibration has lapsed (or drift is present), the device may still report nominal setpoints while the true chamber environment deviates, silently shifting experimental conditions.
Measurement devices can exhibit similar “valid-but-not-comparable” behavior: for luminescence readers, the sensitivity of the PMT depends on wavelength and vendors explicitly provide wavelength-specific calibration, warning that comparability depends on selecting the appropriate calibration wavelength for the assay~\citep{moleculardevices_spectramaxl_userguide}.
In ad hoc automation, these assumptions are rarely modeled, so workflows can complete “successfully” while producing results that are systematically biased or non-comparable across runs.

\begin{tcolorbox}[colback=blue!5!white,colframe=blue!50!black,left=1pt, right=1pt, top=2pt, bottom=2pt]
\looseness=-1
\textbf{Observation \#4:} Reliability challenges arise from both explicit failures (e.g., device errors, sensing failures, or communication timeouts) and implicit failures (e.g., calibration drift or non-comparable measurement settings). Without structured failure handling and calibration-aware state validation, workflows may either halt unpredictably or complete while producing scientifically unreliable results.
\end{tcolorbox}

% \textcolor{red}{check AI.}  
% \rev{
\looseness=-1
Traditional laboratory safety assumes that trained human operators are physically present, with a rich observation space for detecting leaks, abnormal device behavior, unexpected material changes, and other hazards. 
Remote, autonomous laboratories weaken this assumption: scientists interact through APIs, logs, dashboards, and limited telemetry, while physical failures may unfold in real time without direct perception.
This becomes dangerous when autonomous labs execute large batches of AI-generated experiments. 
Explicit failures quickly arise and block downstream runs, while implicit failures such as calibration drift and contamination silently accumulate across many runs. 
Failures that a human might notice and correct locally can therefore propagate through the experiment queue and compromise safety and reliability.
We argue that safety must be enforced deterministically at the system level, across both intra-experiment actions and inter-experiment runs. 
This is necessary not only for large-scale autonomous execution but also to guard against misaligned, erroneous, or malicious scientists.
A centralized lab state is essential: by tracking device availability, configuration, calibration status, environmental conditions, and recent transitions, it enables static validation, runtime checks, state-aware scheduling, and a structured observation space for remote scientists.
% }

\subsubsection{EaC's Execution Layer Addresses These Limitations}
\label{subsubsec:exec-design}

\begin{figure}[t]
\centering
\includegraphics[width=.99\linewidth]{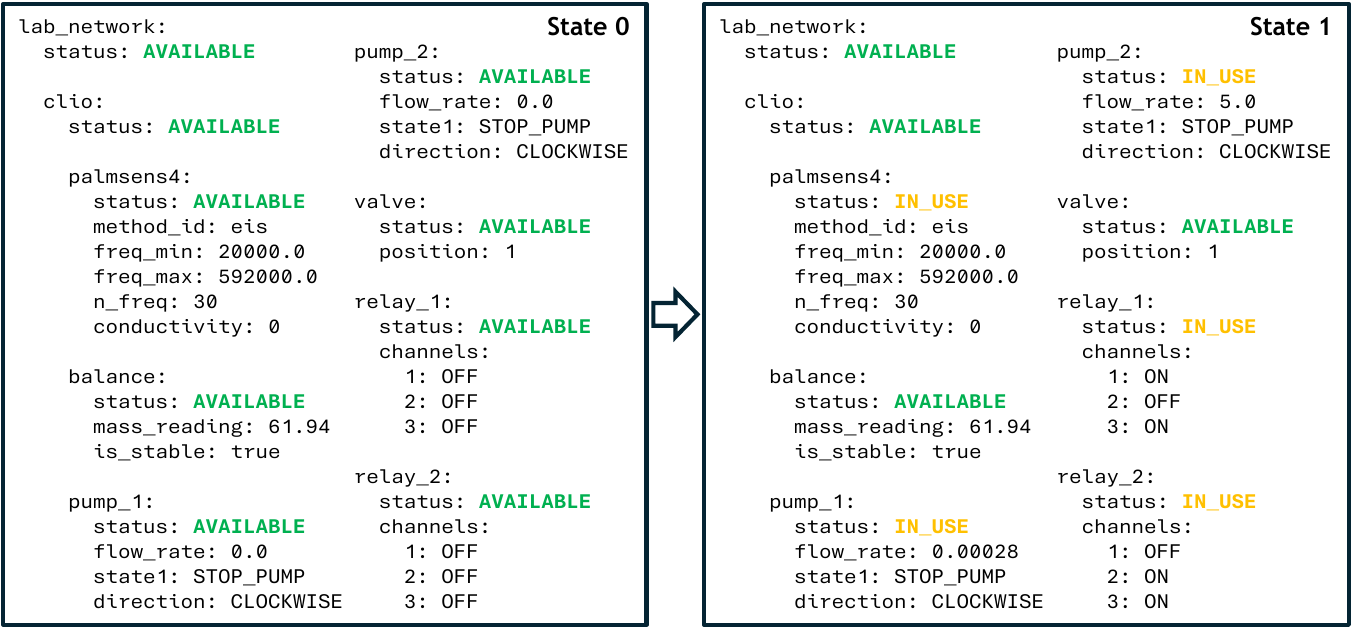}
\caption{The framework maintains a centralized representation of the laboratory state, including device availability, configuration, and runtime status. By tracking state transitions during execution, EaC enables safety checks and prevents conflicting or unsafe operations across devices.}
\label{fig:state-tracking}
\vspace{3mm}
\end{figure}

EaC addresses these limitations by making safety enforcement, fault handling, and state validity first-class concerns of the execution layer.
Before execution, the EaC compiler performs static checks to rule out structurally invalid operations by verifying parameter ranges, capability compatibility, dependency ordering, and device safety envelopes.
During execution, the runtime consults the centralized lab-state model to validate conditions that can only be known at run time, such as device availability, occupancy, operating mode, calibration status, and environmental state.
For example, a request such as \texttt{heater01.heat\_to(}$80^\circ\mathrm{C}$\texttt{)} is allowed only if the heater is free, within its safe operating band, and consistent with its calibration envelope.
Figure~\ref{fig:state-tracking} shows an example centralized lab-state representation and how device states transition during execution.

EaC also treats each instrument as a managed resource with typed operations and explicit failure modes.
For \emph{explicit failures}, the runtime consumes device error channels such as fault codes, exceptions, and communication timeouts, then deterministically transitions the workflow into \emph{pause}, \emph{abort}, or \emph{recover} states.
These transitions are coupled with operation-level idempotence contracts, checkpointed execution logs, and state reconciliation against the lab-state model.
This allows the system to decide whether to retry, compensate, or require human intervention without guessing whether prior steps were partially executed.
For \emph{implicit failures}, EaC encodes validity predicates over device state as part of the execution contract, including calibration provenance, uncertainty bounds, and measurement comparability conditions.
Before issuing a step, the compiler and runtime check that required sensors and measurement pathways are within their validity windows and configured for the intended assay or measurement.
During execution, the runtime compares live telemetry against expected envelopes and escalates when drift, anomalous behavior, or unsafe conditions appear.
By integrating static validation, runtime state checks, and structured recovery, EaC turns physical execution into a state-aware correctness discipline: experiments either complete under validated assumptions or fail explicitly with recoverable state and audit trails.

\subsection{Orchestration Layer}
\label{subsec:orchestration}

The orchestration layer coordinates many validated experiments over shared, stateful laboratory resources, using workflow structure and lab state to improve throughput while reducing unnecessary work.
This layer addresses scalability and efficiency limitations: shared instruments create stateful resource contention, and opaque workflows create waste in time, materials, and human effort, which can be reduced or avoided with better orchestration.

\subsubsection{Scalability}
\label{subsubsec:scalability}

Scalability in laboratory experimentation requires coordinating many workflows over shared, stateful physical resources while preserving safety and correctness.
Unlike compute clusters, instruments are not cleanly time-sliceable: they often require exclusive access, incur nontrivial reconfiguration and stabilization delays, and must remain within calibrated operating envelopes that drift over time.
EaC scales by making these physical constraints explicit and schedulable---i.e., the runtime treats experiments as dependency graphs over \emph{stateful resources}, not just a queue of stateless API calls.

\noindent\textbf{Combinatorial workflow expansion.}
High-throughput experimentation routinely expands a single scientific question into hundreds or thousands of runs (e.g., parallel condition screening and design-of-experiments style sweeps), where throughput hinges on systematically generating, tracking, and executing large batches with consistent semantics~\citep{mennen_hte_review,glauche_doe_hte}.
In such settings, ad hoc scripts often become the bottleneck: they encode brittle assumptions about batching, plate layouts, and instrument availability, and they lack principled mechanisms to reorder or pause safely under contention.

\noindent\textbf{State-dependent reconfiguration overheads.}
Many instruments impose fixed overheads that impact optimal scheduling, but are typically invisible to naive FIFO execution.
For instance, microplate readers often require explicit warm-up periods after power-up (e.g., documented warm-up times and incubation chamber stabilization windows), which makes frequent switching and short jobs inefficient unless the scheduler batches compatible tasks~\citep{biorad_model550_manual,moleculardevices_maxline_manual}.
Similarly, temperature-controlled readers can exhibit long cool-down times (e.g., 12 hours) when switching between assay temperatures~\citep{bmg_assay_stability_cooldown}, so mixing heterogeneous temperature requirements without state-aware batching can destroy utilization.

\noindent\textbf{Multi-tenant contention in shared facilities.}
Cloud labs and shared automation facilities expose large fleets of instruments to many concurrent users. Still, capacity is ultimately bounded by real bottlenecks: scarce devices, maintenance windows, and physical handoffs between workcells. 
Descriptions of modern cloud labs emphasize multi-instrument fleets accessed remotely and operated as a shared service, making multi-tenant contention and scheduling central to throughput~\citep{aws_ecl_story,armer_cloud_labs_perspective,strateos_bioitworld}.

\begin{tcolorbox}[colback=blue!5!white,colframe=blue!50!black,left=1pt, right=1pt, top=2pt, bottom=2pt]
\textbf{Observation \#5:} Scalability challenges arise because laboratory workflows must coordinate exclusive, stateful instruments whose availability and performance depend on physical conditions such as warm-up, cooldown, and maintenance state. Without state-aware scheduling, naive execution leads to resource contention and inefficient utilization.
\end{tcolorbox}

\subsubsection{Efficiency}
\label{subsubsec:efficiency}

Efficiency in empirical scientific discovery comes from rapidly \textit{closing the loop} between high-level scientific hypotheses and validated experiment results and back, which requires \textit{deep visibility into the experiment execution pipeline} in order to minimize manual translation and unnecessary overhead.

\noindent\textbf{Redundant execution across related experiments.}
Exploratory experimental campaigns frequently involve families of protocols that differ only in a small number of parameters (e.g., reagent concentration, incubation duration, or readout gain), while sharing substantial common structure such as plate setup, reagent dispensing, washing, and calibration steps.
In current systems, these protocols are authored and executed as independent scripts or vendor-specific methods, even when their overlap is obvious to human operators. As a result, shared steps are repeatedly revalidated and re-executed, consuming instrument time and reagents without adding new information. Prior work on protocol description languages and cloud laboratories notes that the lack of machine-interpretable structure prevents automated detection of such overlap, forcing conservative treatment of each run as fully independent~\citep{lap_format_repository,bioblocks_autoprotocol}. This redundancy becomes especially costly when scaling to large parameter sweeps or adaptive studies, where dozens or hundreds of near-identical runs are common.

\noindent\textbf{Inefficient queueing and rigid execution order.}
Laboratory execution is heavily shaped by equipment availability, warm-up requirements, and variable run times, yet most current automation systems rely on static or FIFO-style scheduling that ignores these dynamics. Instrument documentation routinely specifies nontrivial setup and stabilization periods (e.g., warm-up, equilibration, or calibration windows), but provides no mechanism for integrating these constraints into a global scheduling policy~\citep{thermofisher_stericycle_manual,nuaire_co2_incubator_manual}. 
Empirical work on cloud-lab optimization notes that, because cloud labs are shared resources, experiments frequently sit in instrument queues, increasing per-experiment cycle time beyond pure device run time~\citep{frisby_protocol_cloudlab, ecl_queuetime_doc, ecl_backlogtime_doc}.
Because protocols are submitted as opaque jobs with fixed step orderings, laboratories cannot reorder compatible steps, batch similar operations, or prioritize jobs based on predicted completion time or downstream dependencies, leading to systematic underutilization of expensive instruments.

\noindent\textbf{Over-conservative waiting under limited observability.}
To hedge against uncertainty, many protocols encode fixed waiting periods that substantially exceed the time required for conditions to stabilize. For instance, incubator and thermal device manuals commonly instruct users to wait long, fixed durations after reaching a nominal setpoint (e.g., ``allow sufficient time for stabilization'') without exposing criteria for determining when stability has actually been achieved~\citep{thermofisher_stericycle_manual,nuaire_co2_incubator_manual}. In practice, these conservative delays accumulate across repeated runs and block downstream resources, particularly in shared facilities. Because existing automation scripts lack access to structured telemetry and cannot express termination conditions beyond fixed timers, early completion cannot be detected and execution remains rigid even when instruments are already within acceptable operating bounds.

\begin{tcolorbox}[colback=blue!5!white,colframe=blue!50!black,left=1pt, right=1pt, top=2pt, bottom=2pt]
\textbf{Observation \#6:} Inefficiency in laboratory workflows often stems from treating experiments as opaque, independent jobs rather than structured pipelines with shared steps and observable state. Without visibility into execution progress and overlap across runs, systems over-wait, re-execute redundant steps, and underutilize instruments.
\end{tcolorbox}

Inefficiency compounds dramatically once AI agents enter the loop. Where a human scientist might propose a handful of follow-up experiments per week, an agent equipped with a hypothesis-generation model can easily emit hundreds of plausible variants per hour. Without an orchestration layer that detects and reuses shared setup and calibration, batches state-compatible jobs, and terminates fixed waits early, the lab quickly becomes the bottleneck and most of the agent's hypotheses will not get tested.
Efficiency therefore stops being an optimization and becomes a hard prerequisite for AI-driven discovery: the value of a faster reasoning model is proportional to how quickly the lab below it can absorb and respond to its proposals.
The same machine-readable workflow structure that enables state-aware scheduling also enables the agent itself to participate in optimization: identifying redundant runs, proposing batch-friendly parameter orderings, and revising plans based on observed instrument utilization. This closes the loop between hypothesis generation and physical throughput.

\subsubsection{How EaC's Orchestration Layer Addresses These Limitations}
\label{subsubsec:orchestration-design}

The EaC orchestration layer schedules compiled workflow DAGs over a centralized lab-state model.
The scheduler reasons about exclusive-use constraints, resource conflicts, inter-step dependencies, and physical handoff constraints.
It also accounts for state transitions with nontrivial latency, including warm-up, cooldown, stabilization, calibration validity, and maintenance windows.
This allows EaC to exploit safe parallelism by running independent workflow branches across distinct eligible devices.
It also enables \emph{state batching}, where tasks with compatible device conditions, such as similar temperature targets or optical configurations, are grouped to amortize reconfiguration overheads.
Because workflows are represented explicitly, the scheduler can reduce unnecessary waiting by advancing steps as soon as required physical conditions are satisfied, rather than relying only on conservative fixed delays.
The same structure allows EaC to avoid redundant work across related experiments by identifying shared setup, calibration, or measurement steps that can be reused when correctness constraints permit.
When contention, delay, or drift occurs, the runtime can deterministically \emph{pause}, \emph{reorder}, or \emph{migrate} tasks to alternative eligible devices based on capability schemas and current lab state.
These decisions preserve the correctness contracts established by the specification and execution layers while improving utilization of shared instruments.
By combining workflow DAGs with state-aware scheduling, EaC makes scalability and efficiency properties of the compiled execution plan and live lab state, rather than fragile consequences of hand-written orchestration scripts.

\section{A Case Study: EaC-enabled Autonomous Battery Lab}
\label{sec:um-case-study}

In this section, we present a case study that illustrates how Experiment-as-Code (EaC) can be integrated into a real-world autonomous scientific laboratory.
We first describe the UM autonomous battery lab, then present its transformation into an EaC-enabled lab, followed by an ionic conductivity experiment that illustrates the benefits of the approach.

\begin{figure}[t]
  \centering
  \includegraphics[width=0.99\linewidth]{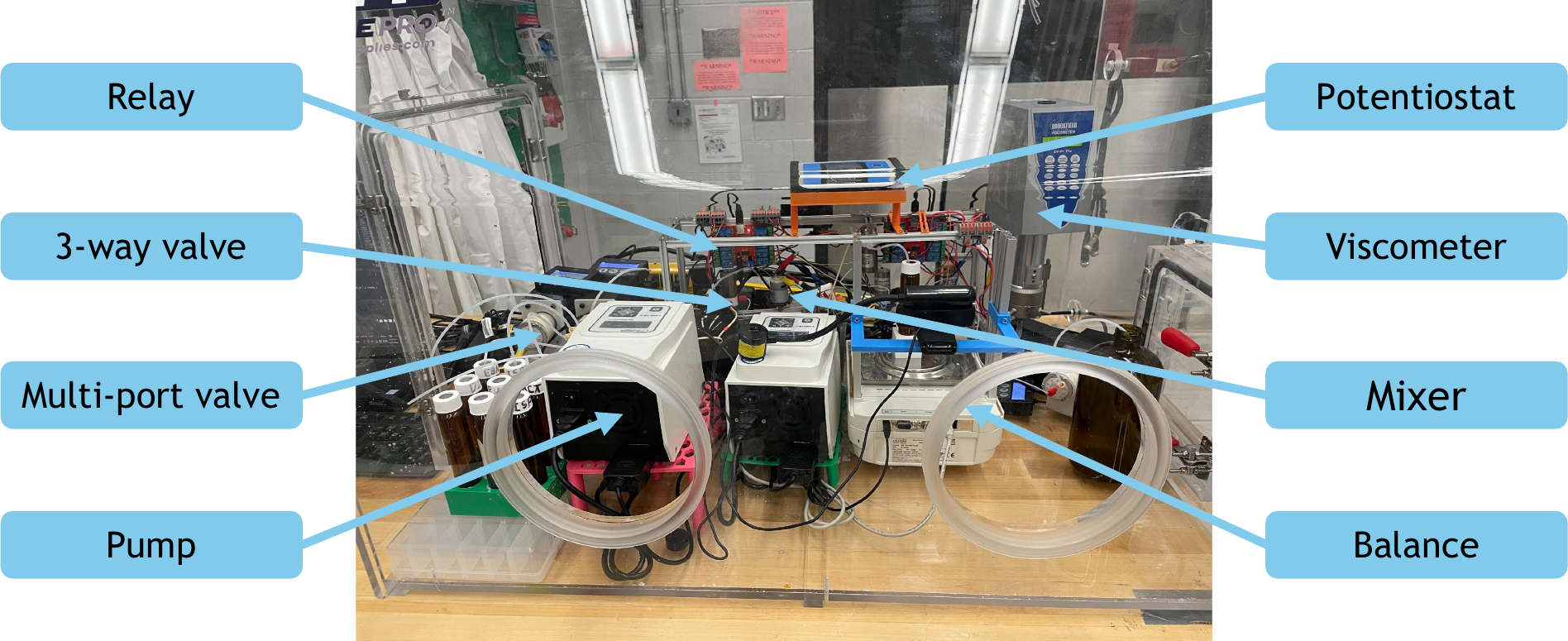}
  \caption{Clio as a custom-built, small-scale proof-of-concept autonomous battery-lab platform.
The platform integrates fluid-handling, measurement, actuation, and control devices within a compact heterogeneous workflow, demonstrating how EaC can standardize devices, coordinate multi-instrument procedures, and maintain shared lab state in a real laboratory deployment.}
  \label{fig:clio_setup}
  \vspace{5mm}
\end{figure}

\subsection{The Battery Lab at UM}

\looseness=-1
We begin with an existing autonomous battery laboratory, which serves as a testbed for introducing EaC.
We chose this setting because it is among the most programmatically controlled laboratory environments available today. 
%While higher levels of automation do supply rich device-level APIs to build against, they also embody the strongest accumulated coupling between scientific intent and vendor-specific control logic: years of hand-tuned scripts, hardware-specific defaults, and undocumented timing assumptions are baked into the existing stack.
%A clean retrofit must therefore reconcile this dense, hard-won imperative logic with a declarative abstraction, rather than enjoy the luxury of a green-field design. If EaC's abstractions can absorb a sophisticated, mature stack without losing fidelity, less automated and more heterogeneous laboratories (which carry less legacy logic to displace) are conceptually downhill.  
The laboratory is designed to accelerate the discovery and characterization of electrolyte formulations for electrochemical systems. 
% Its scientific objective 
The goal is to enable systematic exploration of high-dimensional electrolyte design spaces that are impractical to study through manual experimentation, thereby supporting advances in energy storage and related electrochemical technologies.

\looseness=-1
The platform, \emph{Clio}, shown in Figure~\ref{fig:clio_setup}, integrates multiple experimental instruments under a shared software framework, \emph{ElyteOS}. Clio autonomously prepares liquid electrolyte formulations, executes electrochemical measurements, and characterizes key material properties, including ionic conductivity, density, and viscosity. These properties directly influence electrochemical performance, affecting voltage stability, ion transport efficiency, and degradation mechanisms in batteries and other electrochemical devices.
Electrolyte formulations typically consist of multi-component mixtures of inorganic salts and organic solvents, leading to a large combinatorial search space. To support experimentation at this scale, the lab adopts a modular, flow-through hardware architecture that allows new instruments to be incorporated with minimal physical reconfiguration. The current setup includes high-precision balances, viscometers, potentiostats, peristaltic pumps, multi-port valves, and temperature sensors, all orchestrated through a unified control stack~\citep{dave_autonomous_2022, chen_elyteos_2025}.

Each experimental iteration is automatically logged, version-controlled, and stored in a cloud-accessible database to support reproducibility and retrospective analysis. Experiment selection is currently driven primarily by Bayesian optimization, and prior work has demonstrated tighter learning–experiment loops by coupling robotic execution with differentiable models that encode physical constraints~\citep{zhu_differentiable_2024}.
While these capabilities highlight the lab's increasing degree of autonomy, they also expose a growing gap: as experimental workflows become more complex and adaptive, the mechanisms for specifying, modifying, and reasoning about experiments remain ad hoc and tightly coupled to implementation details. This tension motivates the need for a higher-level, declarative abstraction---setting the stage for Experiment-as-Code transformation.

\subsection{The transformation to an EaC-powered Lab}

% \zy{add lab config here}
% \zy{Add the cfg code snippet in this section.}
\looseness=-1
As a case study, we transform Clio from a script-driven autonomous lab into an Experiment-as-Code (EaC)–enabled platform. While Clio already automates many experimental steps, its original control software relies on low-level, device-specific APIs and hard-coded execution logic. As a result, modifying experiment structure---such as adding new instruments, reordering procedures, or skipping unnecessary measurements---requires invasive changes to core implementations, limiting extensibility and reuse.

To address these limitations, we instantiate the EaC layers described in Section~\ref{sec:eac-core} within Clio, introducing an abstraction layer that declaratively represents both devices and experimental procedures.
At the device level, each instrument is modeled as a managed resource with standardized lifecycle operations (such as create, read, update, and delete) and a unified state contract that captures desired state, observed state, and the reconciliation logic between them.
Concrete device types (e.g., pumps, potentiostats) implement this contract while preserving the uniform interface, so that new instruments can be onboarded without modifying any core orchestration logic.
At the workflow level, we replace hard-coded experiment scripts with declarative specifications: each procedure declares its required resources, actions, parameters, and dependencies, and the resulting DAG is executed by a generic engine that resolves ordering and dispatches to the appropriate device wrappers. The concrete schema of these specifications, and the compilation steps that lower them onto Clio's instruments, follow the layered design described in Section~\ref{sec:eac-core}. 

Using this EaC framework, we implement an experiment to measure the ionic conductivity of aqueous \ce{Li2SO4} electrolytes at varying concentrations (0.43–2.58 mol/kg) under controlled temperature conditions. This experiment requires coordinated control of pumps, valves, and a potentiostat. Unlike the original Clio workflow, which always performs density and viscosity measurements, the EaC specification includes only the procedures required for ionic conductivity measurement. By skipping unnecessary steps, the experiment reduces both execution time and material consumption while preserving correctness. The resulting measurements are consistent with prior experimental studies~\citep{electrical_dissanayake_1988}, demonstrating that the EaC layer can correctly orchestrate physical instruments while offering greater flexibility.

\subsection{Improvements and Preliminary Results}
\begin{figure}[t]
  \centering
  \includegraphics[width=0.7\linewidth]{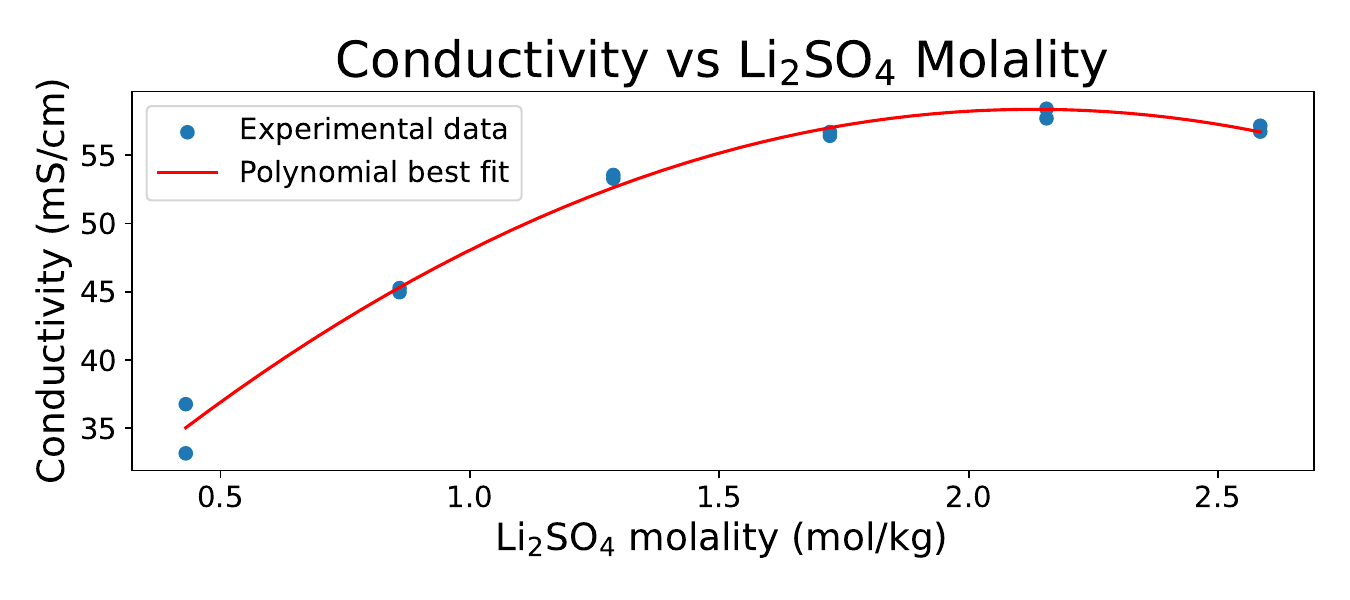}
  \vspace{-3mm}
  \caption{Ionic conductivity measurements for \ce{Li2SO4} aqueous electrolytes at varying concentrations. The experiment identifies 2.15~mol/kg as the highest-ionic conductivity configuration under controlled conditions.}
  \label{fig:experiment_scatter_plot}
  \vspace{3mm}
\end{figure}

To evaluate the practical benefits of EaC, we implemented an ionic conductivity measurement experiment on aqueous \ce{Li2SO4}, a widely studied electrolyte for ion transport analysis in electrochemical systems~\citep{electrical_dissanayake_1988}. The experiment measures ionic conductivity across six concentrations (0.43, 0.86, 1.29, 1.72, 2.15, and 2.58~mol/kg) at a controlled temperature of 298~K.

Executing this workflow in the original non-EaC Clio pipeline would require running the full experimental sequence, including density and viscosity characterization, even when those measurements are unnecessary. In contrast, EaC enables selective specification of experimental intent. We define the experiment through a JSON-based configuration that omits unused procedures, allowing the system to execute only the required steps. This reduces both execution time and material consumption while preserving reproducibility.

The experiment requires coordinated control of multiple devices: a pump transfers electrolyte from designated vials, a multi-port valve routes fluid flow, and a potentiostat performs ionic conductivity measurements. The EaC executor translates the declarative specification into an ordered sequence of operations, including device connection, valve positioning, pump parameter updates, potentiostat configuration, data acquisition, and device teardown. Dependency-aware scheduling ensures correct execution order---for example, preventing measurement reads before fluid dispensing is complete.

As shown in Figure~\ref{fig:experiment_scatter_plot}, the system successfully measures ionic conductivity across all concentrations, identifying 2.15~mol/kg as the highest-performing configuration under the tested conditions. These preliminary results demonstrate that EaC can orchestrate heterogeneous laboratory instruments while enabling cleaner workflow composition and finer-grained experimental control compared to the original tightly coupled pipeline.

\section{Conclusion}
\label{sec:conclusion}

% \rev{
We presented Experiment-as-Code (EaC), a declarative laboratory stack that separates experimental intent from device control to enable standardized specifications, safe execution, and scalable orchestration over shared laboratory resources. 
Our case study demonstrates the feasibility of EaC, showing how it transforms ad-hoc workflows into portable, verifiable experiment programs, improving flexibility and efficiency in autonomous laboratories as a foundation for high-volume AI-driven discovery.

\noindent\textbf{Future directions.}
Experiment-as-Code opens several directions for future work across communities. 
For systems researchers, it raises questions around the abstractions, compilers, schedulers, and state-management mechanisms needed to make physical laboratories programmable, safe, and scalable. 
For AI researchers, it is essential to design agents that can generate experiments through explicit laboratory interfaces, reason over physical and operational constraints, and use structured execution feedback to refine hypotheses, adapt experimental plans, and interpret results.
For domain scientists and laboratory operators, it calls for making experimental procedures explicit: defining step semantics, safety constraints, and validation criteria so that executable protocols faithfully reflect scientific intent.
For instrument vendors, it points to a clear advantage: modular instruments that expose standardized capabilities, telemetry, and control interfaces will be easier to integrate into autonomous laboratories, more reusable across labs and workflows, and increasingly demanded as scientific automation scales.

\noindent
Progress along these directions will require close collaboration across these communities. 
The next breakthrough in AI for Science will not come from a more capable model alone, nor from a more automated instrument alone. 
It will come from the synthesis: a programmable, declarative substrate that lets intelligence and instrumentation co-evolve. 
We view EaC as a step toward that vision and invite the community to build it together.
% }

\bibliographystyle{abbrvnat}
\bibliography{include/pcl, include/danai, include/govAI, include/labs}

\end{document}